\documentstyle[12pt,lathuile,epsfig]{article}
\begin{document}
\title{ 
\bf Search for the Standard Model Higgs boson at LEP
}
\author{
Fabio Cerutti
{\em CERN - EP Division and LNF - INFN} \\
}
\maketitle
\baselineskip=14.5pt
\begin{abstract}
The combined results of the search for the Standard Model Higgs boson 
from the four LEP experiments are given. These results are based on the full
data sample collected by ALEPH, DELPHI, L3 and OPAL at centre-of-mass energies 
up to 209 GeV, corresponding to a total integrated luminosity of about 
2.5 fb$^{-1}$. 
A slight excess of events over the background expectation 
is found at the 2 $\sigma$ level, originating mainly from the ALEPH 4-jet 
channel. This excess is compatible with what expected for the production of 
a SM Higgs boson with a mass of 115.6 GeV/$c^2$. A combined 95$\%$ 
confidence level lower limit of 114.1  GeV/$c^2$ on the mass of the 
Standard Model Higgs boson is derived.
\end{abstract}
\baselineskip=17pt
\newpage
\section{Introduction}
The Standard Model (SM) Higgs boson is expected to be produced at LEP mainly 
via associated production with a Z boson, with small contributions from the W- and 
Z-fusion processes. For this reason the production cross-section is expected 
to have a threshold behaviour at $E_{LEP} \simeq M_{H} + M_{Z}$. For a 115 GeV/$c^2$
Higgs the production cross-section is expected to be about 60 fb at $E_{LEP}$=207 GeV, 
corresponding to 30 signal
events produced in the entire LEP year 2000 data sample.

A  115 GeV/$c^2$ Higgs boson decays mainly into 
$b\bar{b}$ final state (BR=74$\%$), other hadronic final states (gg, $c\bar{c}$, BR=11 $\%$),
$\tau \tau$ (BR=7$\%$) and W$^{*}$W$^{*}$ (BR=7$\%$). 
The final states that are searched for at LEP are:
the 4-jet final state where the Higgs decays into  $b\bar{b}$ and the Z into  $q\bar{q}$,
the missing-energy final state where the Higgs decays into  $b\bar{b}$ and the Z into  $\nu \bar{\nu}$,
the leptonic final state where the Higgs decays into  $b\bar{b}$ and the Z into  $\ell \ell$ (with 
$\ell = \mu$ or e)
and the $\tau\tau$ final state where either the Higgs or the Z decays into $\tau \tau$.

Of these channels the 4-jet has the largest sensitivity followed by the missing-energy, 
the leptonic and the $\tau \tau$.
The b-tag is the main tool used to reject the background coming from other SM processes. 
This variable is very useful in the reduction of the WW, ZZ and $q\bar{q}$ backgrounds. 
Other kinematic variables related to the production and decay of the scalar Higgs boson 
are used in the selection (i.e., the production angle of the selected Higgs candidate, 
the angle between the reconstructed jets, etc.). In addition to these variables the reconstructed Higgs 
boson mass is used to reduce the background that is not compatible with the tested Higgs boson mass 
hypothesis. 

The method adopted by the four LEP experiments consists in computing the 
statistical compatibility of the selected sample (after a loose pre-selection) 
with the two following hypothesis:
the presence of the signal with mass $M_{H}$ (referred to as $s+b$ hypothesis) or 
the compatibility with the expected background (referred to as $b-only$ hypothesis).
This is done by means of a test statistics which takes into account the distributions of
the reconstructed mass, used as the first discriminating variable, and of all the other selection
variables, combined with Neural-Network (NN) or likelihood (L) techniques in the second 
discriminating variable. The test statistics used at LEP is the likelihood-ratio (LR)\cite{AR}, 
the ratios of the likelihood of the selected samples being compatible with $s+b$ and $b-only$
hypothesis. For each tested Higgs boson mass the confidence level of the observed test
statistics being compatible with the two hypotheses are computed: CL$_{s+b}$ and CL$_{b}$.
Signal like results will be characterised by very small values of (1-CL$_{b}$) and by
an average value of CL$_{s+b}$ of 0.5. 
Background like results are characterised by small values of  CL$_{s+b}$ 
(depending on the sensitivity to the tested mass 
hypothesis) and by an average value of CL$_{b}$ of 0.5.

The results reported here are based on about 2.5 fb$^{-1}$ of data collected by the 
four LEP experiments at $E_{LEP}$ ranging from 189 to 209 GeV. 
The most relevant sample is given by the 
542 pb$^{-1}$ of data collected in year 2000 at $E_{LEP} \ge 206$ GeV. 
The combination is based on the results reported in Refs.~\cite{A, D, L2, O} and 
described by the LEP-Higgs working group in \cite{LEP}.
Of these publications only the L3 one is final. 
The ALEPH collaboration has published in Ref.~\cite{A2} its final findings that will
be included in the ultimate LEP combination.

The outline of this paper is the following. In section two the combined LEP results are presented. 
In section three a more detailed discussion of the systematic uncertainties 
is given. In section four the most recent update from the ALEPH collaboration is reported.
Conclusions are drawn in section five.   

\section{LEP combined results}
\begin{figure}[ht]
\begin{center}
 \vspace{1.0cm}
\mbox{\epsfig{file=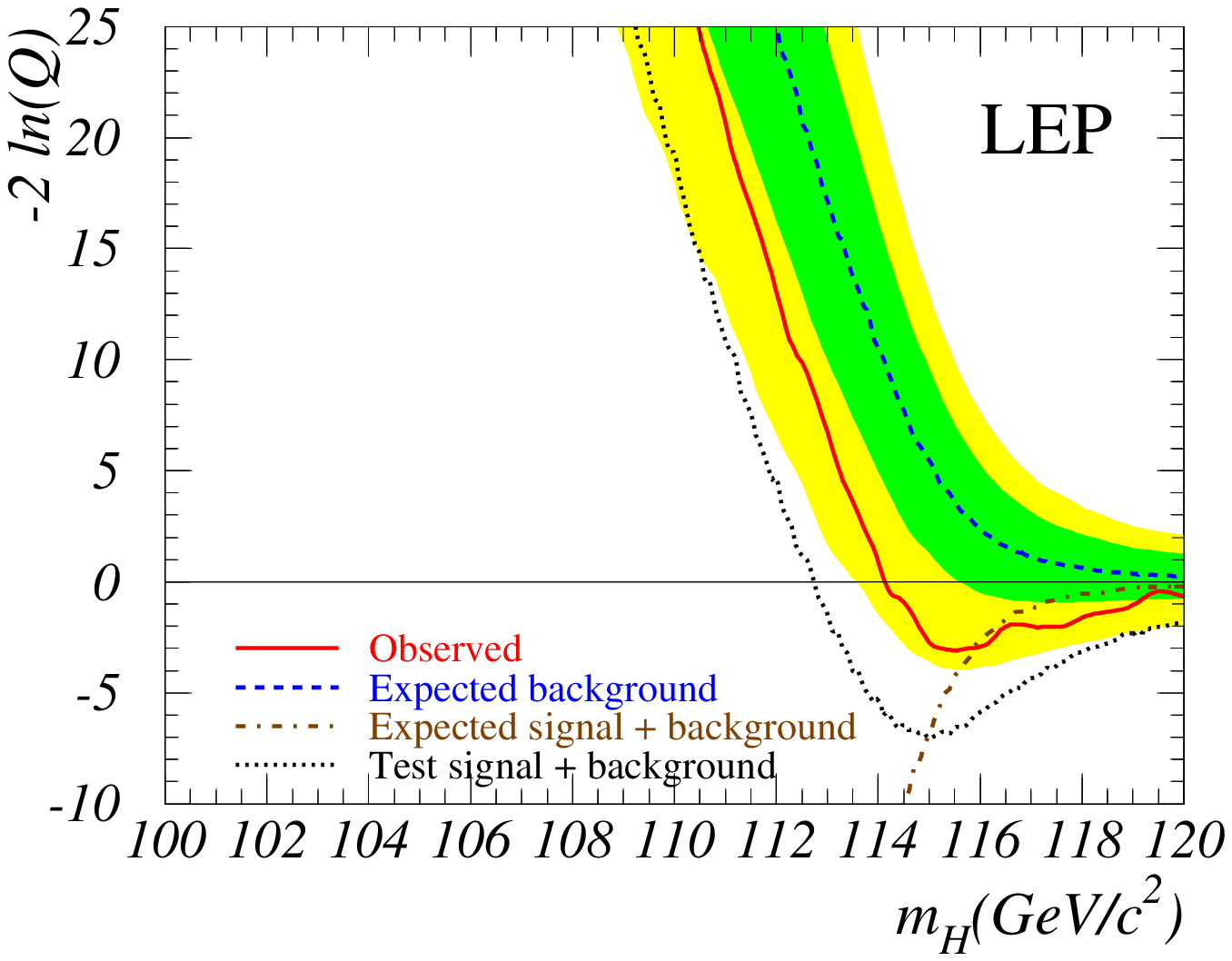,width=16cm,height=13cm} }
 \caption{\it
      Value of the combined test statistics (Likelihood-Ratio) as a function of the tested Higgs boson 
      mass hypothesis. The full line represents the observed result. The dashed line represents the 
expected value for $b-only$ experiments  with 68 (green) and 95 (yellow) $\%$ CL level bands. The value 
expected in case of the presence of a Higgs boson signal is shown by the dash-dotted line. 
The dotted line is obtained with a test where the likelihood ratio distribution has been computed for 
a mixture of the background with a signal with a mass of 115 GeV/$c^{2}$. This test reproduces qualitatively 
the low mass tail (deviation from $b-only$ expectation) of the test statistics observed in the data. 
    \label{f1} }
\end{center}
\end{figure}

The LEP combined LR as a function of the Higgs mass hypothesis is shown in Fig.~\ref{f1}.
The test statistics has a broad minimum around 115.6 GeV/$c^2$ compatible with what is expected 
for an Higgs boson with that mass (CL$_{s+b}$=0.44). The compatibility with the $b-only$ 
hypothesis at that mass is of 3.4$\%$ (i.e., a 2$\sigma$ effect).
The excess is mainly originating from the ALEPH 4-jet channel as can be seen in tables \ref{t1} 
and \ref{t2} 
where the values of CL$_{s+b}$ and (1-CL$_{b}$) are given for the four LEP experiments and for 
the different channels separately. 

\begin{table}[h]
\centering
\caption{ \it Values of the combined  CL$_{s+b}$ and (1-CL$_{b}$) for the four LEP experiments 
and for an Higgs mass hypothesis of 115.6 GeV/$c^2$.
}
\vskip 0.1 in
\begin{tabular}{|l|c|c|} \hline
           &   CL$_{s+b}$ & (1-CL$_{b}$) \\
\hline
\hline
 ALEPH    & 0.94   &      0.002       \\
 DELPHI   & 0.02   &      0.87       \\
 L3       & 0.47   &      0.24       \\
 OPAL     & 0.47   &      0.22       \\
\hline
\end{tabular}
\label{t1}
\end{table}

\begin{table}[h]
\centering
\caption{ \it Values of the combined  CL$_{s+b}$ and (1-CL$_{b}$) for the 
different channels separately and for an Higgs mass hypothesis of 115.6 GeV/$c^2$.
}
\vskip 0.1 in
\begin{tabular}{|l|c|c|} \hline
           &   CL$_{s+b}$ & (1-CL$_{b}$) \\
\hline
\hline
 4-jet            & 0.74   &      0.016       \\
 missing-energy   & 0.26   &      0.40       \\
 all but 4-jet    & 0.19   &      0.34       \\
\hline
\end{tabular}
\label{t2}
\end{table}

From this results a combined 95$\%$ CL lower limit on the Higgs boson mass of 114.1 GeV/$c^2$ can be derived 
to be compared with that of 115.4 GeV/$c^2$ expected in case of absence of the signal.

\section{Systematic uncertainties}

The robustness of the 2$\sigma$ excess observed by the four LEP experiments against 
systematic uncertainties on the expected background has been throughly
investigated. 
Since the excess is mainly originating from the ALEPH 4-jet selection 
only a brief description of the systematic uncertainties related to this channel is given here.
In the threshold mass region ($M_{H} \sim 115$ GeV/C$^2$) the background is dominated by the 
following sources: the ZZ background with one or two Z decaying into $b\bar{b}$, the 
$q\bar{q}$ events with hard gluon emission (with $q$=$b$ or $g \to  b\bar{b}$) 
and the WW background with misidentified b jets.

\begin{figure}[ht]
\begin{center}
 \vspace{1.0cm}
\mbox{\epsfig{file=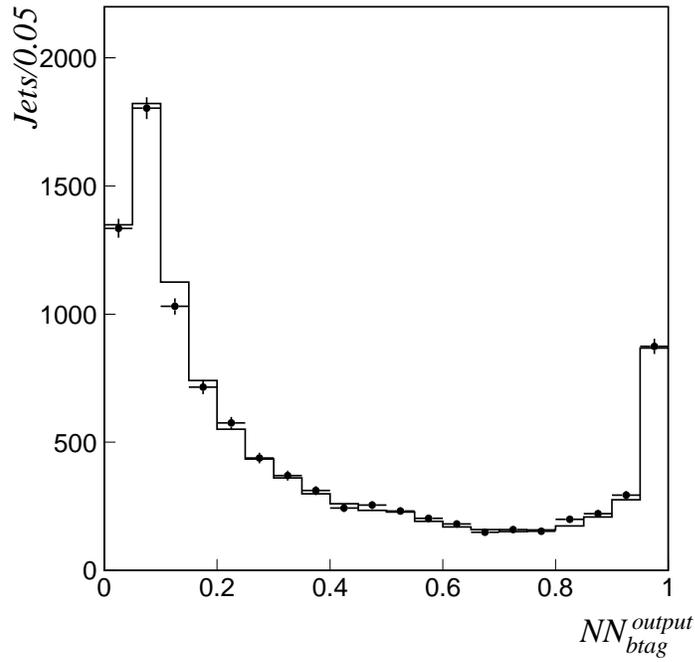,width=10cm,height=10cm} }

\mbox{\epsfig{file=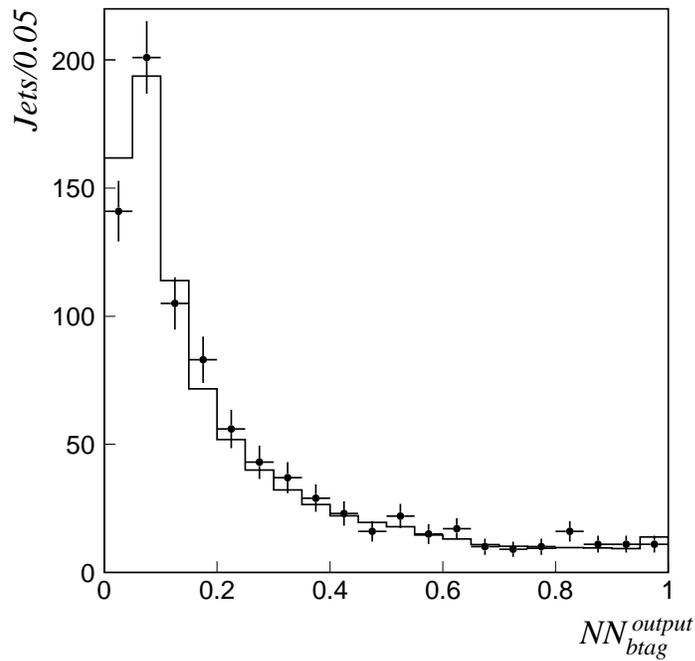,width=10cm,height=10cm} }

 \caption{\it
      Distribution of the jet NN b-tag output for the ALEPH data (points with errors) and for 
      the simulation (open histogram)
      for radiative returns to the Z (upper plot) and semileptonic WW events (lower plot).
    \label{f2} }
\end{center}
\end{figure}

\begin{figure}[ht]
\begin{center}
 \vspace{1.0cm}
\mbox{\epsfig{file=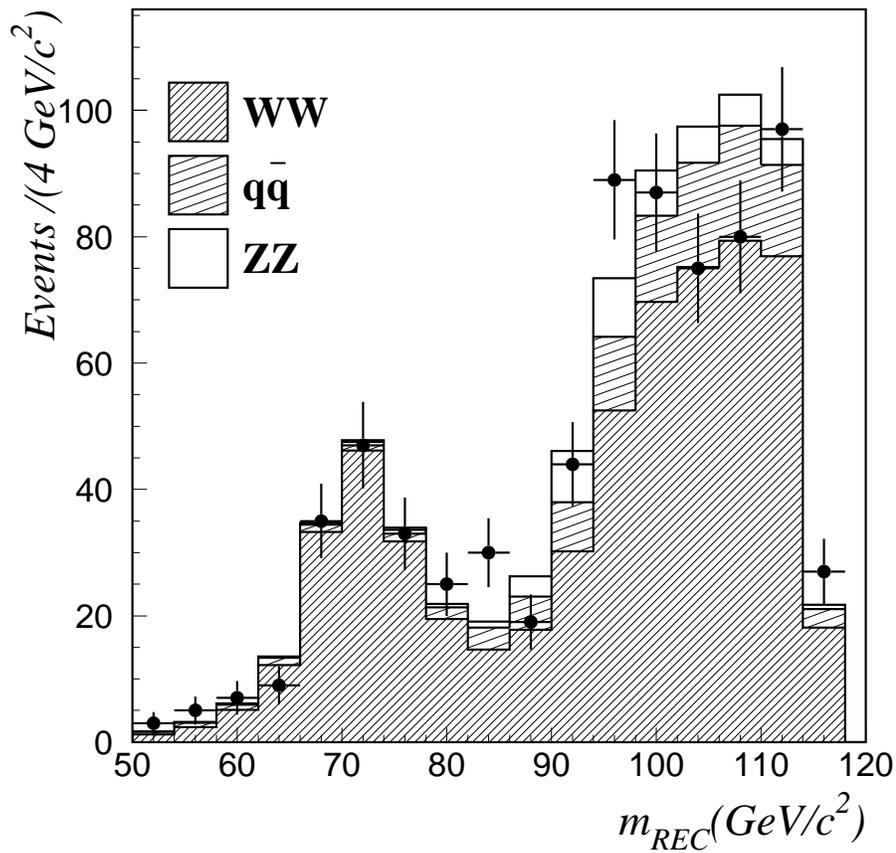,width=13cm,height=13cm} }
 \caption{\it
      Distribution of the reconstructed invariant mass in the ALEPH 4-jet channel for 
     data (points with errors) and for the simulation (histograms), 
     obtained by applying a b-tag veto as explained in the text.
     For the simulation the different background contributions are shown with different shading styles.
    \label{f3} }
\end{center}
\end{figure}

The most relevant selection variables used in the 4-jet channel selection are the b-tag, 
the reconstructed Higgs boson mass and the kinematic-based shape variables.
The b-tag performance in the simulation are calibrated by using the data collected at
the Z peak ($\sim$2.5 pb$^{-1}$ in year 2000). 
Since the decay branching ratios of the Z into heavy and light quarks are well 
known the b-tag efficiency for the $b$ and light-flavor ($udsc$) jets can be measured. 
The simulation parameters have be tuned to reproduce the measured efficiencies and half of
the correction applied to the simulation has been taken as systematic error. The performance of the 
b-tag are then cross checked on high energy data samples containing b jets (radiative returns 
to the Z) or enriched in light flavors (WW events with WW$\to q\bar{q}' \ell \nu$). 
This cross checks are shown for the ALEPH
experiment in Fig.\ref{f2} where the b-tag NN output is compared with the simulation for these 
two samples. The data are in agreement with the simulation within the statistics of the test.
Another important check is related to the understanding of the Higgs mass reconstruction 
close to threshold. This is shown in Fig.\ref{f3}, for a sample 
selected with the same criteria as for the 4-jet selection with the exception of an 
anti b-tag cut applied to reject any possible signal. 
A good agreement between data and simulation is observed for all the mass spectrum.
Other systematic error sources have been investigated by ALEPH: the amount of gluon splitting into
heavy flavors, the description of the quantities used in the second discriminating 
variable, the uncertainty on the correlation existing between the two discriminating variables
and the limited simulation statistics.

The systematic errors estimated by the different experiment on the expected background are already 
included in the values of (1-CL$_{b}$) and CL$_{s+b}$ 
reported in the previous section with the smearing method 
described in ~\cite{CH}. If more conservative techniques are used 
to propagate the systematic errors to the (1-CL$_{b}$) evaluation a reduction of the excess up to 
0.2$\sigma$ is obtained. 

As shown in the next section, the final determination of the systematic uncertainty 
published by ALEPH~\cite{A2} confirms the preliminary value provided for the combination, giving
confidence that the 2$\sigma$ effect is robust against systematic effects.

\section{The final ALEPH publication}
Since the last LEP combination only the ALEPH experiment has published his final
results ~\cite{A2}. 
The main change with respect to the first publication are:
\begin{itemize}
\item the reprocessing of all the data with the final detector calibrations,
\item a more correct treatment of the beam related background,
\item a more accurate estimate of the systematic uncertainty on the expected background.
\end{itemize}

Among these changes only the last two are briefly discussed in this section.
A low polar-angle energy deposit present in one of the three high-purity 
4-jet candidates, candidate $b$ of Ref.\cite{A}, has been interpreted as 
a beam related background. This triggered a study aiming to reject this kind of 
background energy deposits not included in the standard simulation. 
A procedure based on the kinematic fit of the 4-jet event 
with different treatment of the low polar angle clusters has been implemented. 
The impact of this rejection procedure on the background and signal processes
has been studied on simulation samples to which the beam background energy 
deposits, measured on random triggered events, have been added. 
This procedure affects only one data candidate: candidate b of\cite{A}. 
Its reconstructed mass increases from 112.8 to 114.4 GeV/$c^{2}$ and its NN output 
changed from 0.996 to 0.997.

\begin{figure}[ht]
\begin{center}
 \vspace{1.0cm}
\mbox{\epsfig{file=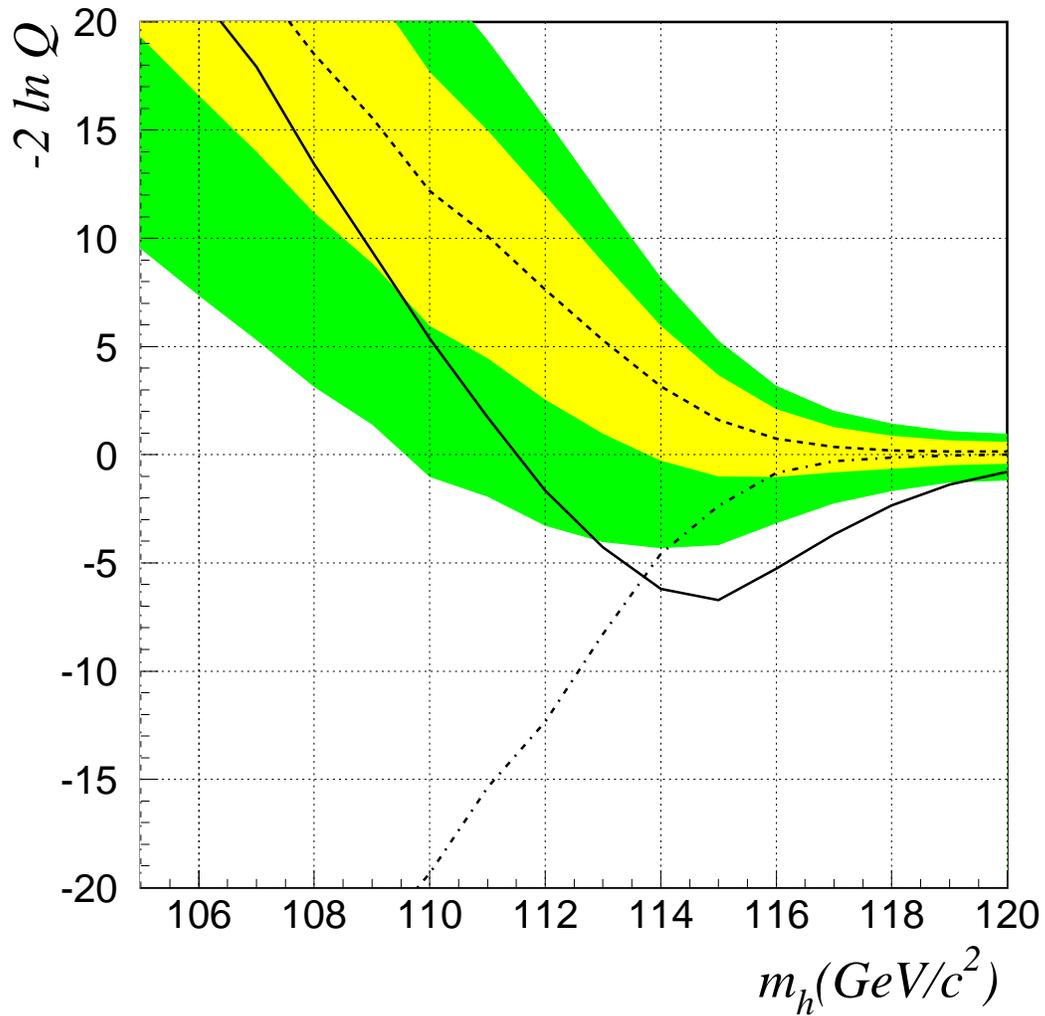,width=13cm,height=13cm} }
 \caption{\it
 Value of the ALEPH test statistics (Likelihood-Ratio) as a function of the tested Higgs boson 
      mass hypothesis. The full line represents the observed result. The dashed line represents the 
expected value for $b-only$ experiments  with 68 (yellow) and 95 (green) $\%$ CL level bands. The value 
expected in case of the presence of a signal is shown by the dash-dotted line.
    \label{f4} }
\end{center}
\end{figure}

The final ALEPH result, given in terms of the LR distribution as a function of the Higgs boson mass 
hypothesis, is shown in Fig.~\ref{f4}. A broad minimum of the test statistics is
observed around $M_{H}$=115 GeV/$c^{2}$. Around this mass value the compatibility with the $b-only$ 
hypothesis is at the level of 2.82$\sigma$ for the NN analysis stream. In the previous ALEPH publication
it was at the 2.96$\sigma$ level. For the alternative analysis stream, the cut-based one, the value of
(1-CL$_{b}$) around the likelihood-ratio minimum corresponds to a significance of 3.04$\sigma$ almost 
unchanged with respect to the 3.06$\sigma$ of the previous ALEPH publication.

With respect to \cite{A} more detailed studied of the systematic uncertainties on the 4-jet background
expectation have been performed.
The systematic errors for the different sources are reported in table \ref{t3} in
terms of changes in the excess significance obtained by varying the background by 
its estimated uncertainty.
The systematic sources investigated are: the statistical uncertainty of the simulation samples, 
the uncertainty on the b-tag variable, the amount of gluon splitting 
into heavy flavours, detector related systematics (like the error on the simulation of 
the jet resolution), the error on the simulation of the other quantities used in the 
second discriminating variables and the error on the strong coupling constant used in the simulation 
(it has an impact in the 4-jet rate in $q \bar{q}$ events).

\begin{table}[h]
\centering
\caption{ \it Impact on the significance, expressed in numbers of $\sigma$,
of the systematic uncertainties on the expected background in the ALEPH 4-jet channel, 
evaluated at $M_{H}$=116 GeV/$c^{2}$.}
\vskip 0.1 in
\begin{tabular}{|l|c|c|} \hline
\hline
Systematic source & $\delta$-significance cut stream &  $\delta$-significance NN stream \\
\hline
 MC statistics             & $\pm$0.11$\sigma$   &   $\pm$0.07$\sigma$         \\
 b-tag                     & $\pm$0.06$\sigma$   &   $\pm$0.08$\sigma$         \\
 gluon splitting           & $\pm$0.04$\sigma$   &   $\pm$0.04$\sigma$         \\
 Jet resolution            & $\pm$0.07$\sigma$   &   $\pm$0.05$\sigma$         \\
 Shape discr. variables    & $\pm$0.03$\sigma$   &   $\pm$0.05$\sigma$         \\
 $\alpha_{s}$ (4-jet rate) & $\pm$0.04$\sigma$   &   $\pm$0.06$\sigma$         \\
\hline
\end{tabular}
\label{t3}
\end{table}

These uncertainties are already included in the confidence levels quoted above by using 
the method described in\cite{CH}. 
If these errors were added in quadrature a total change in the excess significance of less then 
0.2$\sigma$ would be obtained. This confirms the preliminary systematic uncertainty estimated in~\cite{A}.

Since the combined LEP result is driven by the ALEPH 4-jet channel and since the final 
ALEPH publication confirms the findings of the preliminary results, 
the final LEP combination is not expected to change drastically with respect to the 
results reported in section 2. 
Final LEP results are expected by the summer 2002.

\section{Conclusions}
A preliminary combination of the LEP results on the Standard Model Higgs search shows 
a slight deviation from the $b-only$ hypothesis at the 2$\sigma$ level 
(1-CL$_{b} = 3.4 \%$). A closer view to this effect shows that it mainly originates from the 
 ALEPH 4-jet channel. This result is compatible with what expected for a SM Higgs boson with a 
mass of 115.6 GeV/$c^2$ and is robust against systematics uncertainties on the expected background,
as described in the preliminary publications of the four LEP experiments and as 
confirmed by the final ALEPH publication ~\cite{A2}. 
A final clarification on the nature of the observed effect should come from the hadron colliders
TEVATRON and LHC. 

By putting together the indirect constraints, coming from the electroweak fit to the SM 
observable measured at LEP~\cite{LEPEW} (with relevant contribution also from SLD, TEVATRON and NUTEV), 
with the direct search results, the SM Higgs boson mass is constrained in the range between 
114 and 196 GeV/$c^2$ at 95$\%$ CL. This represents an enormous 
improvement on the knowledge of this fundamental observable with
respect to the pre-LEP era. 

\section{Acknowledgements}
I would like to thank all the members of the LEP Higgs working group, Fabiola Gianotti 
and Roberto Tenchini for the help that they gave me in the preparation of this talk. 
I also would like to thank
the organisers of the conference for the high quality of the program, the wonderful
organisation and the beautiful atmosphere in which this conference has been held.  
\newpage

\newpage

\newpage

\newpage

\end{document}